\documentclass[aps,prl,reprint,showpacs,floatfix]{revtex4-1}

\usepackage{graphicx}
\usepackage{bm}
\usepackage{epsfig}
\usepackage{amsmath}
\usepackage{amssymb}
\usepackage{graphicx}
\usepackage{hyperref}  

\newcommand{\CA}{\cal A}
\newcommand{\C}{\cal C}

\newcommand{\al}{\alpha}

\newcommand{\sg}{\sigma}

\renewcommand{\th}{\theta}
\newcommand{\eq}{\begin{eqnarray}}
\newcommand{\eqx}{\end{eqnarray}}
\newcommand{\bal}{\begin{align}}
\newcommand{\eal}{\end{align}}
\newcommand{\ba}{\begin{equation}}
\newcommand{\ea}{\end{equation}}
\newcommand{\fr}{\frac}

\newcommand\n{\nonumber \\}

\begin{document}

\title{Determination of the Odderon amplitude in elastic cross-sections at high energies from scaling and analyticity}

\author{J. A. Velazquez Corral}
\email{jesus.corral@ku.edu}
\affiliation{Department of Physics and Astronomy,
The University of Kansas, Lawrence, KS 66045, USA}

\author{B. G. Giraud}
\email{bertrand.giraud@ipht.fr}
\affiliation{Institut de Physique Th\'eorique, CEA Saclay,
91191 Gif-sur-Yvette, France}

\author{R. Peschanski}
\email{robi.peschanski@ipht.fr}
\affiliation{Institut de Physique Th\'eorique, CEA Saclay,
91191 Gif-sur-Yvette, France}

\author{C. Royon}
\email{christophe.royon@cern.ch}
\affiliation{Department of Physics and Astronomy,
The University of Kansas, Lawrence, KS 66045, USA}

\begin{abstract}
Scaling amplitudes describing $pp$ elastic scattering differential cross-sections in the dip-bump region of momentum transfer at the LHC have been recently 
derived~\cite{scaling}. We check that the same scaling is verified by the $p\bar p$ cross-section at the highest energy of the Tevatron.  Applying the  general "energy to phase" relation for a given  signature, coming from the analiticity properties inherent to the S-Matrix formalism~\cite{chew}, we derive the scaling amplitude with  positive signature (i.e. the Pomeron). Fitting the $pp$ differential cross-sections measured by the TOTEM collaboration leads to some tension with data in the experimental dip observed at moderate momentum transfer. Concentrating the study to the dip/bump region, we are able to determine a contribution of a negative signature amplitude (i.e. the Odderon) leading to  a parameter free prediction for the  $p\bar p$ differential cross-section which  is in agreement with  the D0 data. The extraction of the Odderon amplitude in both modulus and phase is then performed and discussed.
\end{abstract}
\maketitle

\section{1. Introduction}
 High-energy elastic hadron scattering $h_1 h_2 \rightarrow h_1 h_2$ is usually described in terms of the $t$-channel exchange of colorless states carrying no flavor quantum numbers.  One distinguishes two channels depending on  $C$-parity. The leading $C$-even parity channel is dominant and given by the so-called Pomeron exchange which drives the observed rise of total cross-sections with energy. The existence of its $C$-odd partner, called the Odderon exchange, has been proposed long ago~\cite{nicolescu}. One expected consequence of an Odderon would be a difference between proton-proton ($pp$) and proton-antiproton ($p\bar{p}$) elastic scattering even at highest energies~\cite{nicolescu,DL,Bouquet:1975af,Joynson:1975az, Gauron:1992zc}. Until recently, however, the Odderon remained neither experimentally found nor theoretically grounded, even if it has been associated in QCD to a colorless compound of an odd number of gluons, at least three of them~\cite{Nussinov:1975mw,Low:1975sv,Kwiecinski:1980wb}, and since then  has been studied extensively in a perturbative evolution~\cite{Bartels:2019qho,Contreras:2020lrh,Braun:2020gsk,Braun:2023vos} and in bound-state or holographic approaches~\cite{Chen:2021cjr,Hechenberger:2023edv,Ewerz:2003xi}.
 
The experimental status of the Odderon changed decisively when the D0 and TOTEM Collaborations reported the discovery of a C-odd exchange and thus of the Odderon, by comparing $pp$ \cite{TOTEM7,TOTEM8,TOTEM276,TOTEM13} (through a dedicated energy extrapolation) and $p\bar p$~\cite{D0ref} elastic scattering interactions~\cite{TOTEM:2020zzr,Leader:2021zkf}, finding a cross-section difference at energies where other contributions are negligible.

This result motivates a broad theoretical and phenomenological effort to characterize the Odderon contribution, including Regge-based spin-3 oddball and reggeized-exchange models~\cite{Magallanes:2022ddx,Vaso:2026zbo}, Froissaron--Maximal-Odderon parametrisations with spin-flip~\cite{Bence:2020usl,Bence:2021uji}, flat-Odderon~\cite{Petrov:2025ruf} and spin-dependent form-factor~\cite{Benic:2026nnf} descriptions, as well as recent studies of the low-t region~\cite{Luna:2024cbq},  different D0 and TOTEM data comparison and discussion~\cite{Lu:2020wpo,Cui:2022dcm,Donnachie:2022aiq}; recent overviews can be found in Refs.~\cite{Ryskin:2024qpq,Royon:2026joy}.\\

On top of the Odderon discovery, the TOTEM Collaboration has measured the $pp$ elastic differential cross-section $d\sigma/dt$ at $\sqrt{s} = 2.76$, $7$, $8$ and $13$~TeV as a function of $|t|$. For different $\sqrt{s}$ values, and in the transfer momentum range of the so-called dip-bump region, the cross-sections do not vary in an independent way: they exhibit a pattern in which the curves are shifted, hinting at the presence of a scaling variable, combining the $s$ and $t$ variables in data~\cite{scaling}. Building on these findings, a largely model-independent line of evidence leads to a successful description of data across all the energies measured by TOTEM~\cite{scaling} using a scaling amplitude. Additional scalings have been found in elastic $d\sigma/dt$ data~\cite{Csorgo:2019ewn}, and the real-extended Bialas--Bzdak (ReBB) model; the Odderon has also been considered through a Barger-Phillips type extrapolation. They led to additional evidence for the Odderon contribution ~\cite{Csorgo:2020wmw,Szanyi:2022ezh,Szanyi:2022qgx,Csorgo:2023rzm,Goncalves:2018nsp}. 

However, until now, no clear, model-independent determination of the Odderon amplitude had been found. It is the purpose of this Letter to present such a determination in the context of the scaling~\cite{scaling}  properties of the high energy elastic amplitudes and of the analytic S-matrix formalism~\cite{chew}.

For this sake, we shall follow the scaling law established in Ref.~\cite{scaling}, defined across all the energies measured by TOTEM, which we refer to as the scaling framework. In this Letter, we study the signature content of these scaling amplitudes within the general analyticity properties of the  $S$-matrix~\cite{chew}, without committing a priori to any specific  model. The energy-to-phase relation imposed by the formalism leads to the determination of phases, distinguishing the positive-signature (Pomeron) and negative-signature (Odderon) components and fixing their relative phase. This phase relations is the key new ingredient for the determination of the Odderon amplitude. Working  within the scaling framework~\cite{scaling}, and using the appropriate energy-to-phase relations, we are able to extract the corresponding Odderon amplitude from the study of $pp$ data~\cite{TOTEM7,TOTEM8,TOTEM276,TOTEM13} at the LHC and obtain predictions for the $p\bar p$ elastic cross-section~\cite{D0ref} at the Tevatron.

The structure of this Letter is the following. In Sections~2 to~4, we introduce the scaling proton-proton amplitudes and recast them in the  $S$-matrix formalism, separating the contributions of fixed and positive signature and deriving the energy-to-phase relation characterizing the analyticity properties of the S-matrix framework. In Section~5, we work on experimental data and extract the Odderon amplitude and then the $p\bar{p}$ one from the implications of these rules  for the Tevatron and the LHC ranges of energy: Fitting the scaling behaviour of $pp$ elastic scattering at the LHC, we obtain a parameter-free prediction for the $p\bar{p}$ cross-section, and finally  extract the Odderon amplitude in both modulus and phase. We present a discussion in Section~6 and conclude in Section~7.

\section{2. Scaling proton-proton scattering amplitudes}
\label{1}

The scaling framework defined in Ref.\cite{scaling} describing scaling properties of proton-proton elastic scattering at LHC energies have been empirically revealed, based on the elastic differential cross-sections measured by the TOTEM  collaboration  \cite{TOTEM7,TOTEM8,TOTEM276,TOTEM13}. 

 Consider the  $pp$  differential cross-section  
\eq
   \fr {d\sg}{dt}(s,t)\ &=& \  \vert{\CA}(s,t)\vert^2\ ,
\label{proton}
\eqx
where  $s$ is the positive center-of-mass energy 
squared and $t$ is the momentum transfer squared.
${\CA}(s,t)$ is the spin-averaged proton-proton elastic  
scattering amplitude. For 
convenience, the formula Eq.\eqref
{proton} is normalized in a dimensionless  way such that the 
amplitude  is dimensioned to an energy to the power ($-2$). 

 The scaling 
 properties observed in Ref.~\cite{scaling} are such that a 
conveniently rescaled cross section
 appears to be depending on one variable only, $ t^{**} $, combining $s $ and $ t$. This reads
\ba
s^{-\al}\  {\fr {d\sg}{dt}}(s,t) =  
f \left([s^{\nu} |t|]^{.72} \equiv t^{**}\right]\ ,
\label{crossscalingzero}
\ea
where $\al/2=0.1525$ and $\nu= 
0.0903.$

Hence, the  property Eq.\eqref{crossscalingzero} means that 
the 
amplitude ${\CA}(s,t)$ verifies a scaling property characterized by the two scaling exponents  $\al$ and $\nu.$ Indeed one can  rewrite \eqref{crossscalingzero} in terms of a scaling amplitude
\ba
 {\CA}(s,t) = s^{\al/2}
F \left[s^{\nu}\ t \right]\ ,
\label{crossscaling}
\ea 
where $F$ is a function of a single variable. A possible
 overall phase, not present in the cross-section, will be discussed later on. Eq.\eqref{crossscaling} shows that, up to the overall  rescaling factor $s^{-\al/2},$ the amplitude
${\CA}(s,t)$ may be written using the scaling variable
\ba
\tau\ \equiv \ |t| \times {(s/ { {\rm TeV^2}})^{\nu} } \ . 
 \label{scalingvariable}
\ea

In Ref.\cite{scaling}, one  finds a concrete realization of the scaling \eqref{crossscaling}
with a good fit of the measured cross-sections 
by the TOTEM collaboration \cite{TOTEM7,TOTEM8,TOTEM276,TOTEM13} in the dip-bump region of 
momentum transfer. The
amplitude  verifying the scaling writes
\eq
{\cal A} &=& e^{i\th} 
({\cal A}_{1}+ {\cal A}_{2})\n
{\cal A}_1 &=& i \ N_1  \ e^{-B_1 |t|}\ =   i \ N_1  \ e^{-B_1^0 \tau}\n 
{\cal A}_2 &=& i  \ N_2  \ e^{-B_2 |t| +i\varphi}\  = i  \ N_2  \ e^{-B_2^0 \tau +i\varphi}, 
\label{amplifit}
\eqx
where 
\begin{eqnarray}
N_{1,2} &=& N_{1,2}^0 (s/ { {\rm TeV^2}})^{\fr \al 2} \\
B_{1,2} &=& B_{1,2}^0 (s/ { {\rm TeV^2}})^{\nu}
\end{eqnarray}
where the parameters $N_k^0,B_k^0,  {k=1,2} $ and the  phase
 $\varphi$  were phenomenologically fitted  to data~\cite{scaling}. $\th$ is chosen to be the known phase due to the Coulomb interference in the very 
forward direction ($\theta=-0.09$ rad). The fit values of the parameters are provided together with the other fits of our study in Table~\ref{parameters}, section 5.2.

\section{ \label{2} 3. Scaling amplitudes with fixed signature}

The problem we want to introduce and study is how to describe a scaling 
amplitude such as
Eq.\eqref{amplifit} in the S-matrix formalism with its  constraints related to its analyticity properties~\cite{chew}. Note, to be clear, that we do not rely on any specific phenomenology such as for instance Regge models. Only general analyticity properties of an elastic amplitude are involved.

Let us formulate these analyticity properties. In the S-matrix  framework, the amplitude ${\CA}$  can be 
written  as the integral over a sum of two components in the complex plane of the $t$-channel partial wave amplitudes.
\ba
   {\CA}_\eta(s,t)\ =\ \fr 1{2\pi i}\int_{\C} {s^{l} +\eta\ (-s)^{l} \over \sin \pi l}\
   \ a_\eta (l,t), \ ,
\label{regge} 
\ea
where $\eta =\pm 1$ is called the signature. 

The partial wave amplitudes $a_\eta(l,t)$ are the analytic 
continuations    in 
the crossed  $t-$channel two-body process, with definite signature. The contour $\C$ is turning around the poles corresponding to even or odd integer zeroes of $\sin \pi l$ depending on  the signature $\eta = \pm 1$, since the poles of $\sin \pi l$ of opposite $\mp$ parity are canceled by the numerator zeroes. It can eventually be deformed and curled around other singularities in the complex $l$-plane such as Regge poles and cuts, well-known in Regge phenomenology. 	In our study we will make no assumption about  singularities other than those due to  $\sin \pi l$ in Eq.~\eqref{regge} and stick to the general analytic S-matrix formalism leading to Eq.\eqref{regge}.

We note that the integrand of \eqref{regge} can also be written
\eq
\!\!\! {s^{l} +\eta\ (-s)^{l} \over \sin \pi l}
&=& \ \fr {(e^{-i {\pi\over 2}} s)^l}{\sin (\pi l/2)}\quad {\rm if}\ \eta=+1 \n
&=& \ i\ \fr {(e^{-i {\pi\over 2}} s)^l}{\cos (\pi l/2)}\quad {\rm if}\ \eta=-1\ .
\label{reggephase}
\eqx
exhibiting the typical $\pi/2$ phase ratio between positive and negative signature integrand factors.
Hence, as shown by Eq.\eqref{reggephase}, the analytic  S-matrix formalism 
 imposes a  phase induced by  the $s$ dependence, apart from dynamical phases of the partial waves $a_\eta(l,t).$ One thus obtains an ``energy-to-phase" relationship
through the substitution of variables
\ba
s\ \Rightarrow\ \sg\  =\ e{^{-i {\pi}/ 2}}s \ .
\label{relation}
\ea

In conclusion of this section, there are quite stringent consequences of 
Eq.\eqref{relation} since the  energy dependence of the 
amplitudes $ {\CA}_\eta$ depends only on the complex variable $\sg.$ Hence, there exists
a precise energy-to-phase relationship, Eq.\eqref{relation}, predicted by the analytic S-matrix formalism independently of any  model.
Moreover,  for a negative  signature amplitude, there appears  a phase shift of $\pi/2$ with respect to the positive signature amplitude.

\section{ \label{sec:3}4. Scaling amplitudes with positive signature}

We shall now apply the general formalism depicted in Section (2) to the scaling amplitudes proposed in Ref.\cite{scaling}, cf. Eqs.\eqref{amplifit}. For this, we operate the substitution Eq.\eqref{relation} to the  energy variable dependence of these amplitudes

We are thus led to superimpose the energy-to-phase relation \eqref{relation} to the parameterization \eqref{amplifit}.  
\eq
{\cal A}_+ &=& \  {\cal \tilde A}_{1}+ 
{ \cal \tilde A}_{2}\n
{\cal \tilde A}_{1} &=&
{\tilde N}^0_1 \ 
\left({\sg \over {\rm TeV^2}}\right)^{\al /2} 
 e^{{-\tilde B_1}\tau}\n
{\cal \tilde A}_{2} &=&
{\tilde N}^0_2 \ 
\left({\sg \over {\rm TeV^2}}\right)^{ \al /2} 
 e^{{-\tilde B_2}\tau +i\tilde\varphi}
\label{amplifitregge}
\eqx
where we  use a modified notation with tildes in order to make explicit the positive signature parameterization, and distinguish it from the initial one \eqref{amplifit}. Note  that the parameters ${\tilde N}_k, {\tilde B_k}$, and the phase $\tilde\varphi$  are obtained from  fits to data, while the overall phase, contrary to \eqref{amplifit} where it was chosen as the Coulomb phase, is constrained by the energy-to-phase relation \eqref{relation} applied to the scaling factor, namely $s^{-\al/2} \to \ \sigma^{-\al/2}.$

An explicit expression  is thus
\eq
 \tilde {\cal A}_{1} &=& \ i\ {\tilde N}_1  \ 
\left({s \over {\rm TeV^2}}\right)^{\al /2} e^{-i {\pi}\al/ 
4}\ \times \n
&\times&\ e^{\left\{{-\tilde B_1}\tau
\ \left(\cos  {\pi \nu} - i\sin {\pi \nu}\right)\right\}}\ ,\n
 \tilde {\cal A}_{2} &=& \ i\ {\tilde N}_2  \ 
\left({s \over {\rm TeV^2}}\right)^{\al /2} e^{-i {\pi}\al/ 
4}\ \times \n
&\times&\ e^{\left\{{-\tilde B_2}\tau
\ \left(\cos  {\pi \nu} - i\sin {\pi \nu}\right)\right\}+i\tilde\varphi}\ ,\label{amplifitredefined}
\eqx
where we use the scaling variable $\tau$ defined by \eqref{scalingvariable}.
We recover  the same  form as the initial components of \eqref{amplifit} by the parameter relations $B_k = {\tilde B_k}\cos
\fr {\pi \nu}2$ and $N_k = \tilde N_k.$ 
 Indeed one  can write
\eq
{\cal A}_+ = e ^{-i\frac{\pi\al}4}
\left({\cal A}_{1}
e^{i B_1 \tau \tan (\fr {\pi \nu}2)}
+ {\cal A}_2\ 
 e^{i \left(B_2 \tau \tan (\fr {\pi \nu}2)\right)}\right)\n
\eqx
or, after factorization of an overall phase, 
\eq  
\ e^{i \left( B_1 \tau\  \tan \fr {\pi \nu}2 -\fr{\pi\al}4\right)}
\left({\cal  A}_{1} +\ {\cal  A}_{2}\
e^{(B_2-B_1)\ \tau \tan \fr {\pi \nu}2}\right)
\ ,\n
\label{amplifitfactorized}
\eqx
where the ${\cal  A}_{k=1,2} $ have the same parametric form as in the initial  scaling  
formula, Eq.\eqref{amplifit}.

The net result of implementing the analytic S-matrix constraints amounts  to adding  $\tau$-dependent phases to  the fixed  ones, namely
\eq
\th \ &\Rightarrow& \ B_1 \tau\ \tan (\pi \nu/2 ) -\fr{\pi\al}4\ \equiv\ \th_\tau,\n
\varphi \ &\Rightarrow& \ \tilde\varphi+\ \left(B_2-B_1\right) \tau\ \tan  (\pi \nu/2)\ \equiv\  \varphi_\tau\ .
\label{regge phases}
\eqx
Therefore, we may rewrite Eq.\eqref{amplifitfactorized} as
\eq
 {\cal A}\ &=&\   e ^{i  \th}
\  \left(\ {\cal A}_{1}\  
+ \ {\cal A}_2\ \right)\n
 {\cal A}_+ &=&\   e ^{i \th_\tau}
\left(\ {\cal A}_{1}
+ \ e^{i[\varphi_\tau-\varphi]}{\cal A}_2\ \right)\ .
\label{ppamplitude}
\eqx

Note that, due to the analyticity constraint for signature $\eta=+1,$  we find in the first line of Eqs.\eqref{regge phases}
$\th_\tau \ne \th,$  which was the phase chosen from the Coulomb interference parameter in Eqs.\eqref{amplifit}. This will have an influence in the determination of the Odderon amplitude.

\section{\label{5} 5. Extracting the Odderon amplitude}

\subsection{5.1. Scaling behavior at the Tevatron and LHC}
Let us first test if the same scaling described in Eq.~\eqref{crossscalingzero} is valid for $p\bar p$ elastic scattering differential cross-section as for $pp.$ We thus apply this scaling to the $d\sigma/dt$ data at $\sqrt{s} = 1.96$~TeV measured by the D0 Collaboration at the Fermilab Tevatron~\cite{D0ref}. Following the  functional form for the scaling variable $t^{**}$ introduced in Ref.~\cite{scaling}, and adopting the same scaling exponents,  we obtain the scaling transformation given in~Eq.\eqref{crossscalingzero}.

\begin{figure}[h]
\scalebox{0.43}
{\includegraphics*{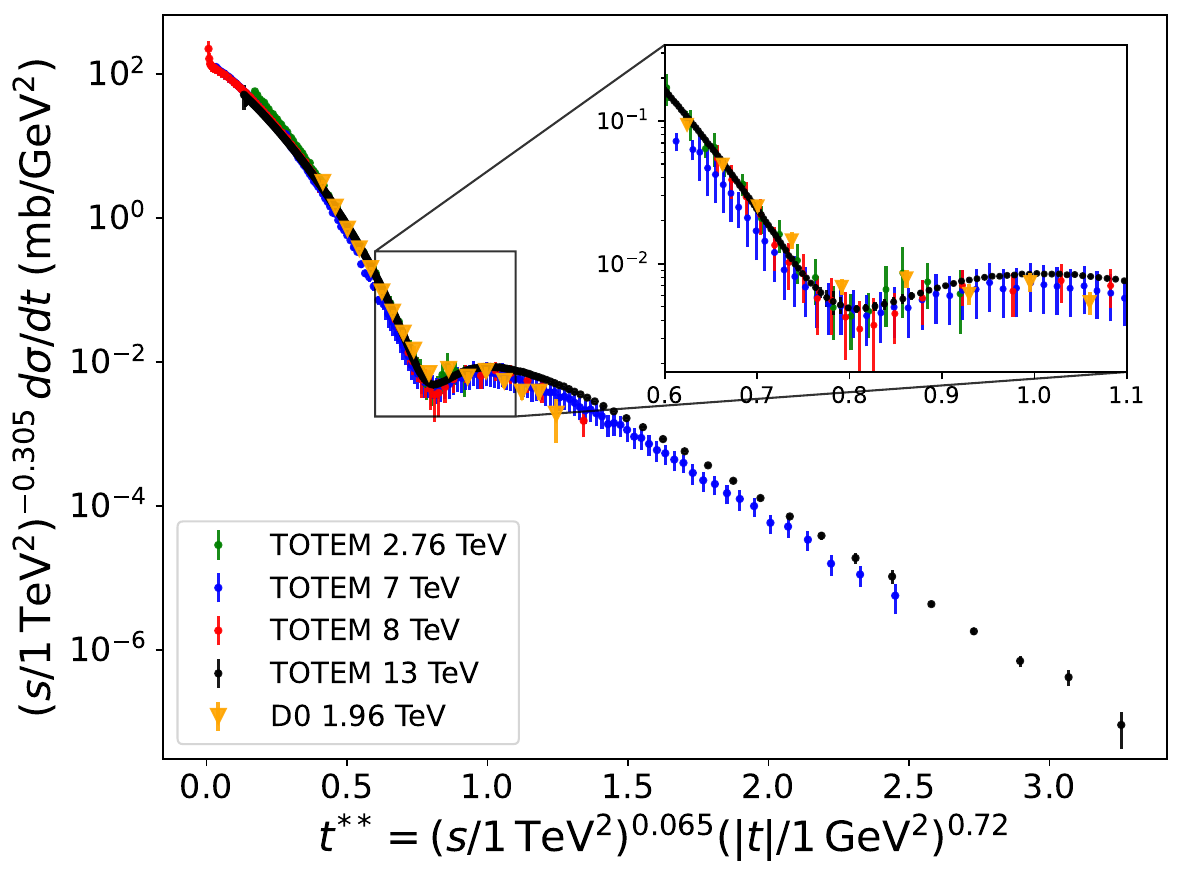}}
\caption{Scaled elastic $p\bar{p}$ and $pp$ differential cross-sections as a function of the scaling variable $t^{**}$. The results show the consistency of the scaling laws across different collision systems and energies. We also check that the D0 elastic data follow the same scaling as the TOTEM data. However, the zoom region in the 
dip/bump region shows a small difference between the $p\bar{p}$ (yellow) and  
$pp $ data points (green, blue, red and black corresponding to different center-of-mass energies).}
\label{fig333}
\end{figure}

\begin{figure*}[!htbp] 
    \centering
    \begin{minipage}{0.48\textwidth}
        \centering
        \includegraphics[width=\linewidth]{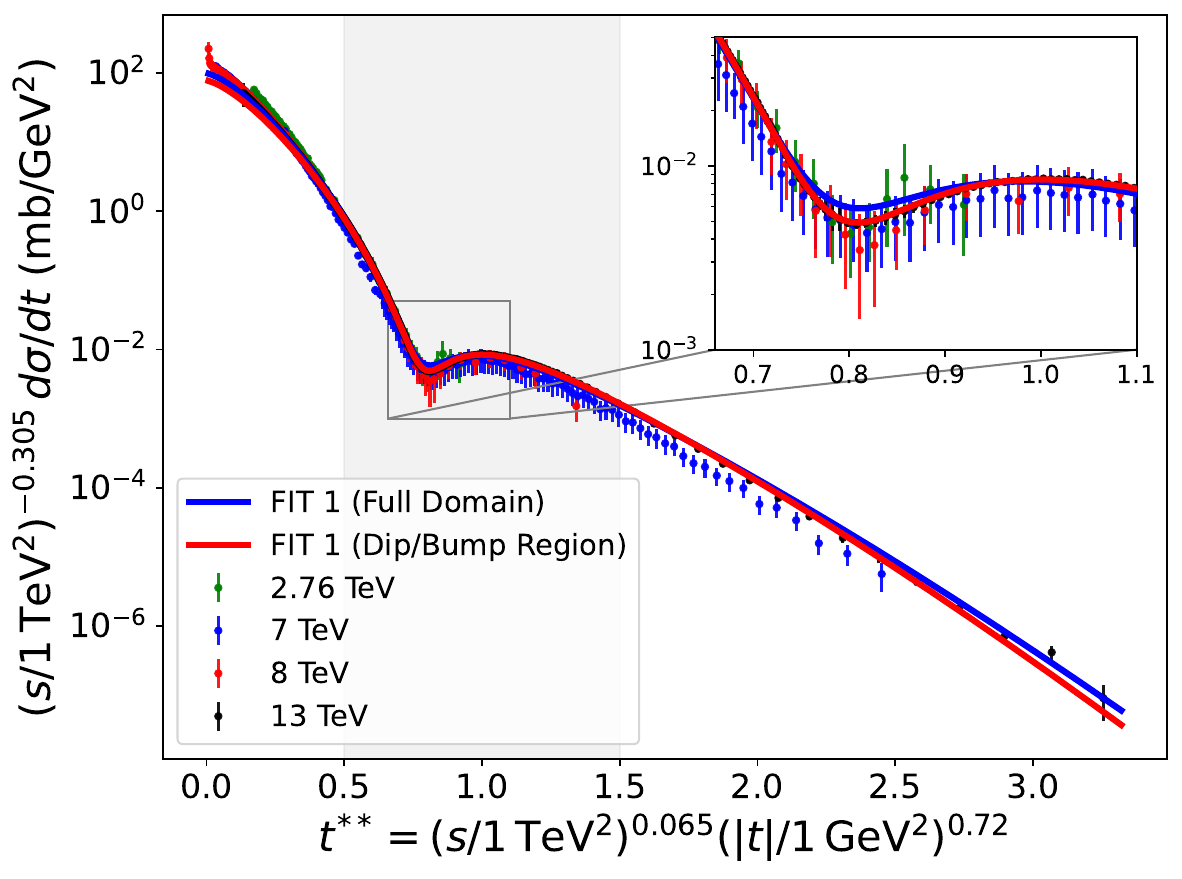}
    \end{minipage}
    \hfill
    \begin{minipage}{0.48\textwidth}
        \centering
        \includegraphics[width=\linewidth]{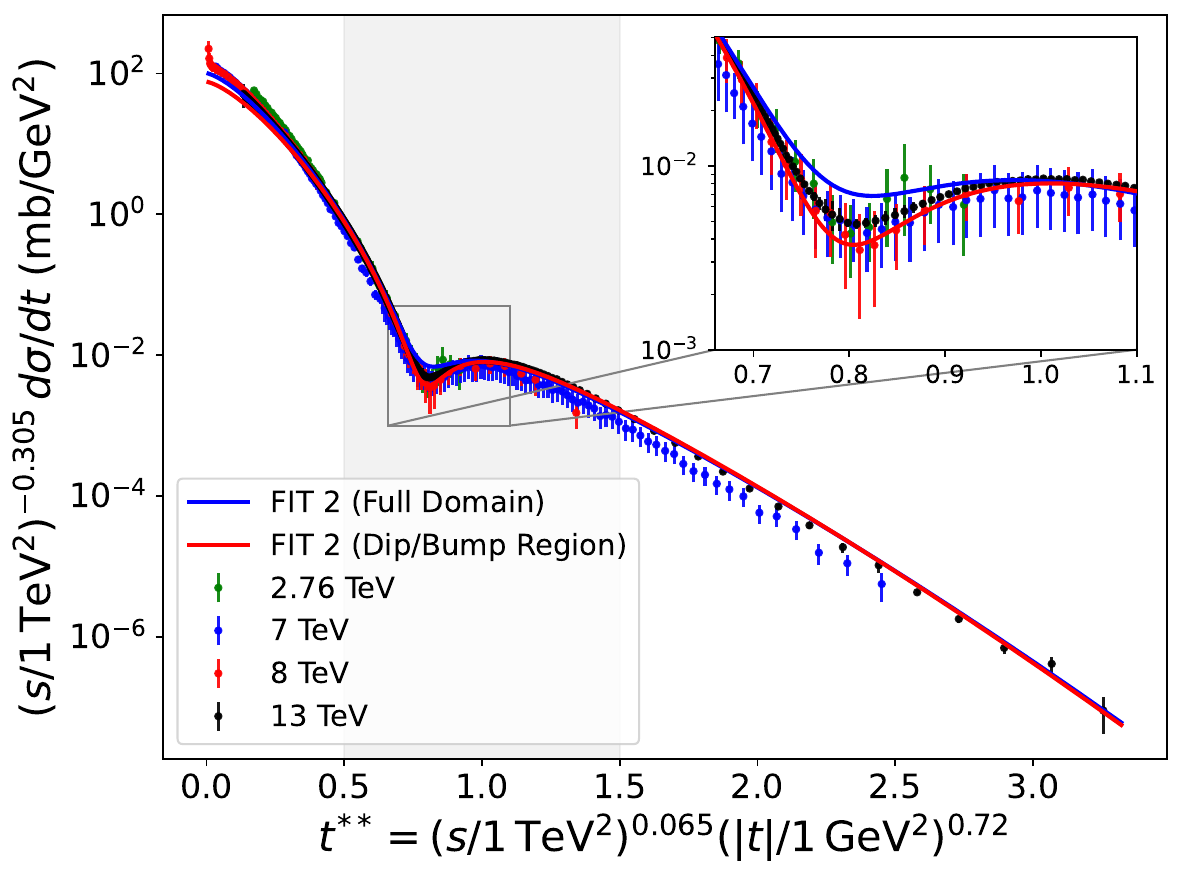}
    \end{minipage}
   \caption{\textbf{Left:} Fits of the full amplitude ($A_{pp}$) to the scaled TOTEM elastic scattering data, $(d\sigma/dt) (s/\text{TeV}^2)^{-0.305}$, as a function of $t^{**}$ using either all data (blue solid line) or data in the dip/bump region only (red solid line). \textbf{Right:} Similar fits using the positive signature amplitude ($A_{+}$) only. The grey area shows the dip/bump region only. We also show the zoomed results in the $0.65 < t^{**} < 1.1$ GeV$^2$ region to display the differences in the dip/bump region between both fits. The difference is larger at high $t^{**}$ as expected.}
    \label{fig:fits}
\end{figure*}
First, the momentum transfer $|t|$ is mapped to the scaling variable $t^{**}$ using:
\begin{equation}
    t^{**} = \left( \frac{s}{1\,\text{TeV}^2} \right)^{0.065} \left( \frac{|t|}{1\,\text{GeV}^2} \right)^{0.72}.
\end{equation}

Subsequently, the scaled differential cross-section is computed as:
\begin{equation}
    \left( \frac{d\sigma}{dt} \right)_{\text{scaled}} = \left( \frac{s}{1\,\text{TeV}^2} \right)^{-\alpha} \frac{d\sigma}{dt},
\end{equation}
where $\alpha = 0.305$ is the same scaling exponent determined in Ref.~\cite{scaling}. This procedure allows for a direct comparison between the $p\bar{p}$ data and the $pp$ results within the same kinematic scaling plane. Fig.~\ref{fig333} shows both scaled elastic $p\bar{p}$ and $pp$ differential cross-sections as a function of the scaling variable $t^{**}$, which indicates that $p \bar{p}$ elastic data follow the same scaling as the $pp$ ones.

However it is important to note that  the zoomed region in  Fig.~\ref{fig333} in the 
dip/bump region shows a small difference between the $p\bar{p}$  and the   
$pp $ data points. The zoomed region shows a rather flat behavior of the scaled elastic $p\bar{p}$ cross section  in contrast with the dip and bump characteristic of the scaled elastic $pp$ cross section. It leads alkready to some preliminary hints of a possible negative signature component.
Despite these small differences both $pp$ and $p \bar{p}$ shows the same scaling laws. 

We will thus perform the fits using the $t^{**}$ variable in the following to extract the Odderon.

\subsection{5.2. Fits to TOTEM elastic data}

We recall the physical  differential elastic cross section formula \eqref{proton}
\eq
\frac{d\sigma}{dt} = \ |{\cal A}(s,t)|^2\,,
\label{xc}
\eqx
and  the double exponential parameterization for the elastic amplitude \eqref{amplifit}
\eq
{\cal A}(s,t) = \left( {\cal A}_1(s,t) + {\cal A}_2(s,t) \right) e^{i\theta}
\label{ampli1}
\eqx
where,
\eq
{\cal A}_1(s,t) &= i  N_1(s)\ e^{-B_1 (s)|t|}\n
{\cal A}_2(s,t) &=i  N_2(s)\ e^{-B_2 (s)|t|+i\phi}
\label{ampli2}
\eqx

\begin{table*}[htbp]
\label{table1}
\centering
\begin{tabular}{|c||c|c||c|c|}
\hline
Parameter                   & FIT 1        & FIT 1 & FIT 2 & FIT 2 \\
 & all data & dip/bump region & all data & dip/bump region \\
\hline
$N^0_1$/$N^{0+}_1$ ($\sqrt{mb} GeV^{-1}$)  & 10.39 $\pm$ 0.04   &   9.36 $\pm$ 0.06  & 10.21 $\pm$ 0.04    & 9.23 $\pm$ 0.06   \\
$N^0_2$/$N^{0+}_2$ ($\sqrt{mb} GeV^{-1}$)  &   0.49 $\pm$ 0.01   &  0.59 $\pm$ 0.01   &  0.49 $\pm$ 0.01  &  0.53 $\pm$ 0.01   \\
$B^0_1$/$B^{0+}_1$ ($GeV^{-2}$)  &   5.78 $\pm$ 0.01   &   5.42 $\pm$ 0.02   &  5.91 $\pm$ 0.02   &  5.38 $\pm$ 0.03    \\
$B^0_2$/$B^{0+}_2$ ($GeV^{-2}$)  &   1.44 $\pm$ 0.01   &   1.52 $\pm$ 0.02    &   1.44 $\pm$ 0.01   &    1.46 $\pm$ 0.02   \\
$\phi$/$\phi^+$ (rad) &  $\pm$ 2.68 $\pm$ 0.01  &  $\pm$ 2.77 $\pm$ 0.01    &  -1.96 $\pm$ 0.01   &  -2.88 $\pm$ 0.002  \\
$\theta$ (rad) & -0.09   &   -0.09  &  -0.09  & -0.09 \\ 
\hline
$\chi^2$/ndof                                  & 5.923    &  1.1 &  5.809 & 1.12 \\
\hline
data points                                  & 597    &  270 & 597  & 270 \\
\hline
\end{tabular}
\caption{Parameters for FIT 1 and FIT 2 to all TOTEM $p \bar{p}$ data and to data in the dip and bump region only.}
\label{parameters}

\end{table*}

This is where we use the scaling properties of elastic data in the fitting procedure, and
the differential cross section is fitted as a function of $t^{**}$ with the $s$ dependence given by  scaling.

\eq
\frac{d\sigma}{dt} =  [N_{1}(s)]^2 e^{-2B_{1}(s)|t|}+[N_{2}(s)]^2 e^{-2B_{2}(s)|t|} \n
+ 2N_{1}(s)N_{2}(s) e^{-[B_{1}(s)+B_{2}(s)]|t|}\cos{\phi}
\label{fit1}
\eqx
where the functions $N_{1}(s),N_{2}(s), B_{1}(s)$ and $B_{2}(s)$ are defined as:
\eq
N_{1,2}(s) &= N_{1,2}^{0} \left( \frac{s}{\text{TeV}^{2}} \right)^{\alpha/2}\n
B_{1,2}(s) &= B_{1,2}^{0} \left( \frac{s}{\text{TeV}^{2}} \right)^{\gamma/2}
\label{ampli3}
\eqx

From the scaling results~\cite{scaling},  the values of $\alpha = 0.305$ and $\nu=\gamma/2 \sim 0.0903$ are imposed. The model contains six free parameters: $N_{1}^{0}, N_{2}^{0}, B_{1}^{0}, B_{2}^{0}, \phi$, and $\theta$. We fix $\theta  = 0.09\ \rm{rad}$ to satisfy $\rho (t = 0) = Re(A)/Im(A) = 0.10$ at high energies, as measured by the TOTEM
experiment~\cite{scaling}.

We then fit the data of the TOTEM Collaboration for all $\sqrt{s}$ in the scaling plane, 
\eq
  \frac{d\sigma}{dt} \left( \frac{s}{\text{TeV}^{2}} \right)^{-0.305} 
  \eqx
 as a function of
 \eq
    \n   t^{**} = \left( \frac{s}{\text{TeV}^{2}} \right)^{0.065} \left( \frac{|t|}{\text{GeV}^{2}} \right)^{0.072}
\eqx
using the expression in Eq.~\eqref{fit1}. Experimental uncertainties are treated as uncorrelated. We refer to this double-exponential fitted function as \textbf{FIT 1}. We also consider fitting data in the dip/bump region only (defined as $0.5 < t^{**} < 1.5$) or in the full $t^{**}$ domain.\\

The first fitting method uses the full amplitude $A_{pp}$, a double-exponential function (FIT~1) given by Eq.~\eqref{fit1}, fitted to scaled elastic $pp$ data measured by TOTEM at 2.76, 7, 8, and 13 TeV \cite{TOTEM7,TOTEM8,TOTEM276,TOTEM13} as a function of $t^{**}$. The fit to all data (blue solid line in Fig.~\ref{fig:fits}, left) yields $\chi^{2}/\mathrm{ndof} = 5.92$ (597 data points), and the data in the dip/bump region only, defined as $0.5 < t^{**} < 1.5$ (red solid line in Fig.~\ref{fig:fits}, red), which yields $\chi^{2}/\mathrm{ndof} = 1.1$ (270 data points), in order to extract the odderon and $p\bar{p}$ elastic amplitudes in the following section.\\

We consider a second series of fits, denoted  \textbf{FIT 2}, following the discussion in Section 4 and  we use the amplitude based on the positive signature contribution only. Consequently, the resulting fit must be performed as before, but replacing the values of $\th $ and $\phi $ by the expressions  of  $\th_\tau $ and $\phi_\tau $ as shown in Eq.~\eqref{regge phases}. To account for the fact that the fitted parameters for the standard amplitude ${\cal A}$ and the positive signature amplitude ${\cal A}_+$ are independent, we introduce a superscript $+$ to denote the parameters belonging to ${\cal A}_+$  for FIT 2 (i.e., $N_1^+, N_2^+, B_1^+, B_2^+, \phi^+$).

The two distinct amplitudes are defined as
\eq
{\cal A}(s,t) &=& \left( N_1(s)e^{-B_1(s)|t|} \right. \n
&& \left. +\ N_2(s)e^{-B_2(s)|t|}e^{i\phi} \right) e^{i\th}\  \n
{\cal A}_{+}(s,t) &=& \left( N_1^+(s)e^{-B_1^+(s)|t|} \right. \n 
&& \left. +\ N_2^+(s)e^{-B_2^+(s)|t|}e^{i\phi_\tau} \right) e^{i\th_\tau}\ 
\label{ampli_independent}
\eqx
where the S-matrix phases for the positive signature component are constructed strictly using the parameters of the positive fit
\eq
\th_\tau &=& B_1^+ |t|s^{0.09} \tan(0.045\pi) -\fr{\pi\al}4\ , \n
\phi_\tau &=& \phi^+ +\ \left(B_2^+ - B_1^+\right) |t|s^{0.09} \tan(0.045\pi)\
\label{regge_phases_independent}
\eqx

The differential cross-section is thus fitted using the following expression:

\eq
\frac{d\sigma_{+}}{dt} =  [N_1^+(s)]^2 e^{-2B_{1}(s)|t|}+[N_2^+(s)]^2 e^{-2B_{2}(s)|t|} \n
+ 2N_1^+(s)N_2^+(s) e^{-B_1^+(s)+B_2^+(s)]|t|}\cos{\phi_{\tau}}
\label{fit1bis}
\eqx
The parameters are refitted for the positive amplitude only. Figure~\ref{fig:fits}, right, displays the second fitting method results. It fails to describe the dip bump region, present in the elastic $pp$ data.  This is  expected since this fit considers only the pomeron contribution to the elastic cross-section. The parameters for both fits are shown in Table~\ref{parameters}. It is worth noticing that the fits do not depend on the sign of $\phi$ since $\phi$ only appears in a cosine for the cross section formulae. We will see in the next section that the sign will appear relevant for the prediction of the Odderon cross section.

\begin{figure*}[htbp] 
    \centering
    \begin{minipage}{0.48\textwidth}
        \centering
        \includegraphics[width=\linewidth]{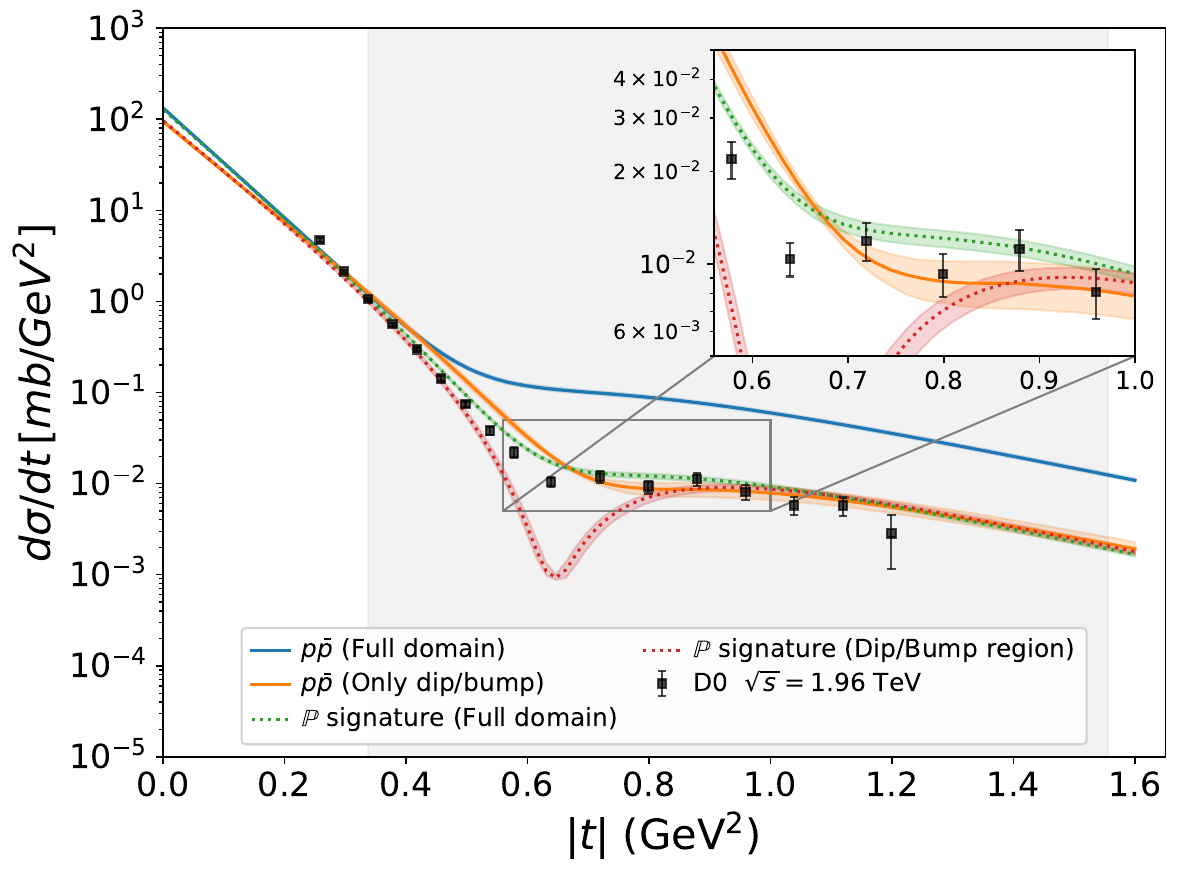}
    \end{minipage}
    \hfill
    \begin{minipage}{0.48\textwidth}
        \centering
        \includegraphics[width=\linewidth]{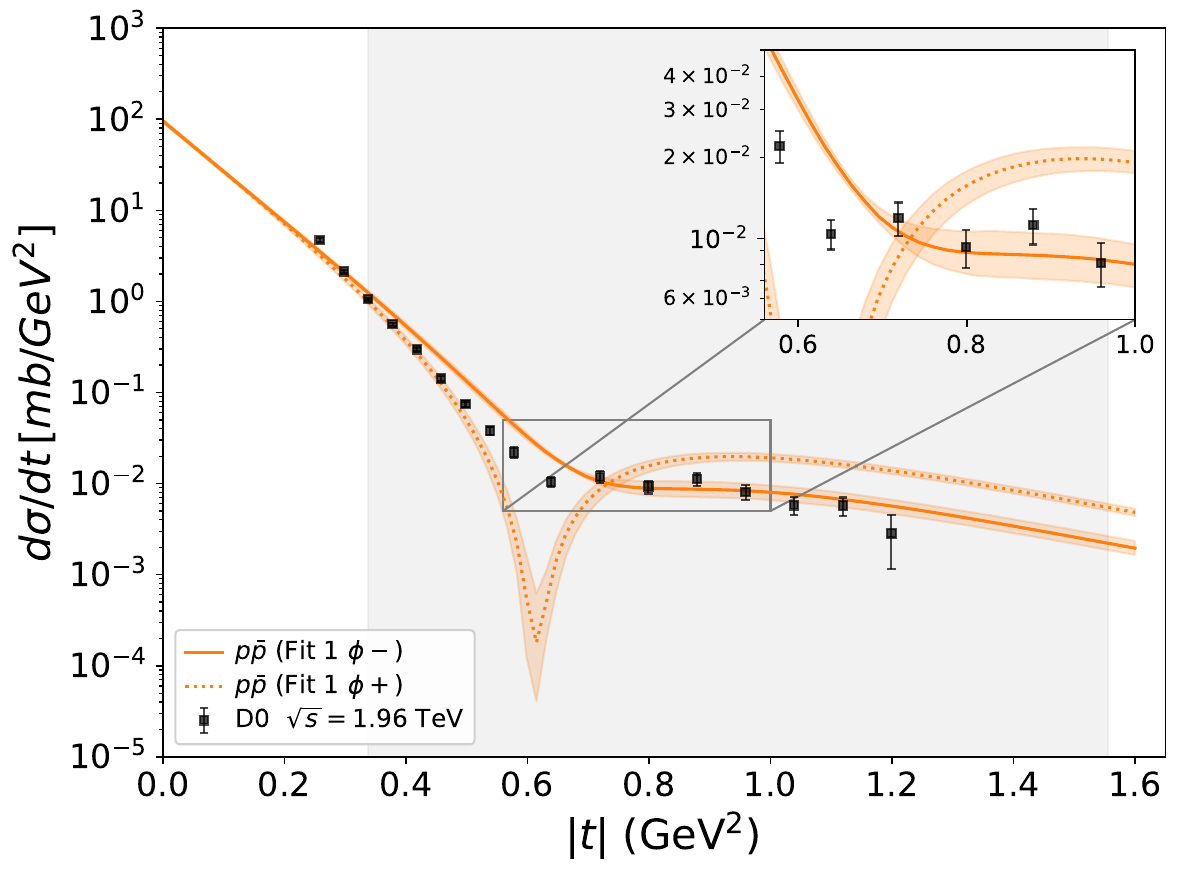}
    \end{minipage}
    
    \caption{\textbf{Left:} Predictions for the $p\bar{p}$ elastic differential cross-section at $\sqrt{s} = 1.96$ TeV derived from FIT 1 and FIT 2 to TOTEM $pp$ data using scaling. The D0 data (black points) are compared to the cross sections computed using $|A_{p\bar p}|^2$  (solid lines) and the pomeron signature $|A_{+}|^2$ (dotted lines). We also show the grey domain that corresponds to the dip/bump region ($0.5 < t^{**} < 1.5$ GeV$^2$ or $0.34 < |t| < 1.56$ for $\sqrt{s}=1.96$ TeV). The shaded bands indicate the $1\sigma$ propagated uncertainties. 
    \textbf{Right:} Impact of the free sign of the $\phi$ from Fit 1. Both figures are zoomed  in the dip and bump region.}
    \label{fig:d0_both}
\end{figure*}

\subsection{5.3. Formulas for the extraction of the Odderon and $p\bar{p}$ amplitude}

Using the relation ${\cal A}_- = {\cal A} - {\cal A}_+$, and the fits described in the previous section using the $ pp$ amplitude ${\cal A}$ and the positive signature one (or Pomeron) ${\cal A}_+$, we can extract the negative signature (or Odderon) ${\cal A}_-$  amplitude by subtraction. It reads
\eq
{\cal A}_-(s,t) &=& \left[ N_1(s)e^{-B_1(s)|t|}e^{i\th} \right. \n
&& \left. -\ N_1^+(s)e^{-B_1^+(s)|t|}e^{i\th_\tau} \right] \n
&& +\ i \left[ N_2(s)e^{-B_2(s)|t|}e^{i(\th+\phi)} \right. \n 
&& \left. -\ N_2^+(s)e^{-B_2^+(s)|t|}e^{i(\th_\tau+\phi_\tau)} \right]\ 
\label{aminus_independent}
\eqx
We then are able to obtain the $p \bar{p}$ amplitude through the relation ${\cal A}^{p\bar p} = 2{\cal A}_+ - {\cal A},$ namely 
\eq
{\cal A}^{p\bar p}(s,t) &=&  \left[ 2 N_1^+(s)e^{-B_1^+(s)|t|}e^{i\th_\tau} \right. \n
&& \left. -\ N_1(s)e^{-B_1(s)|t|}e^{i\th} \right] \n
&& +\  \left[ 2 N_2^+(s)e^{-B_2^+(s)|t|}e^{i(\th_\tau+\phi_\tau)} \right. \n
&& \left. -\ N_2(s)e^{-B_2(s)|t|}e^{i(\th+\phi)} \right]\! 
\label{appbar_independent}
\eqx

\begin{figure*}[htbp] 
    \centering
    \begin{minipage}{0.48\textwidth}
        \centering
        \includegraphics[width=\linewidth]{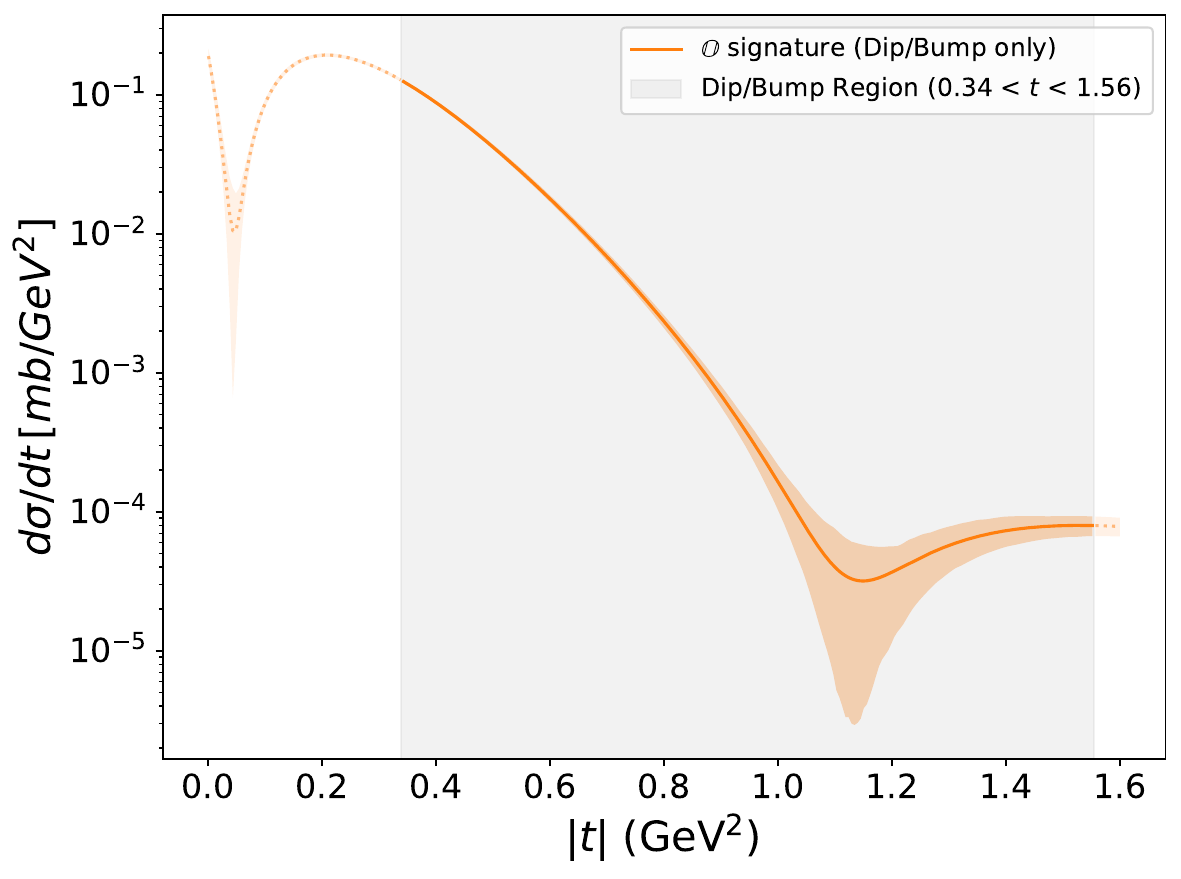}
    \end{minipage}%
    \hfill
    \begin{minipage}{0.48\textwidth}
        \centering
        \includegraphics[width=\linewidth]{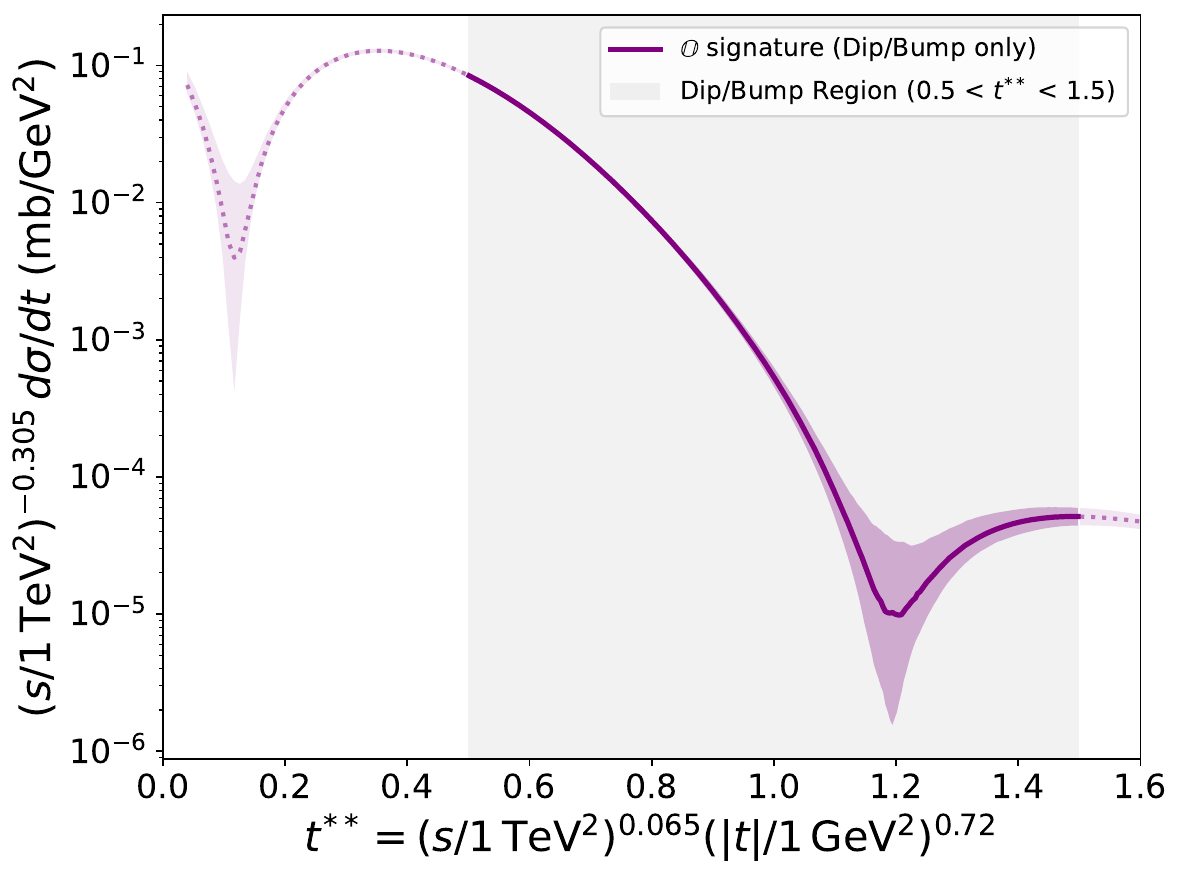}
    \end{minipage}
    
    \caption{\textbf{Left:} Predictions for the Odderon signature on the elastic differential cross section at $\sqrt{s} = 1.96$ TeV (D0 energy) as a function of $|t|$. The curves are derived from the TOTEM $pp$ fits via the scaling framework using the $A_{-}$ amplitude, with shaded bands showing the $1\sigma$ propagated uncertainties. \textbf{Right:} Same prediction plotted as a function of the scaling variable $t^{**}$ within the TOTEM dip-bump region.}
    \label{fig:d0_bothb}
\end{figure*}

To compute the differential cross-sections, we define the purely real magnitudes of the exponential components of both the standard and positive-signature amplitudes, which simplifies the resulting expressions and their implementation in our calculations. For the standard amplitude ${\cal A}$:
\eq
T_1(s,t) &=& N_1(s)\ e^{-B_1(s)|t|}\  \n
T_2(s,t) &=& N_2(s)\ e^{-B_2(s)|t|}\ 
\label{t1t2_standard}
\eqx
and for the positive-signature amplitude ${\cal A}_+$:
\eq
T_1^+(s,t) &=& N_1^+(s)\ e^{-B_1^+(s)|t|}\  \n
T_2^+(s,t) &=& N_2^+(s)\ e^{-B_2^+(s)|t|}\ 
\label{t1t2_positive}
\eqx
Taking the squared modulus of the negative signature amplitude $|{\cal A}_-|^2 = |{\cal A} - {\cal A}_+|^2$ from Eq.~\eqref{aminus_independent}, and applying Euler's formula to resolve the absolute squares and cross-terms, we obtain the differential cross-section
\eq
\fr{d\sigma_-}{dt} &=& T_1^2(s,t) + [T_1^+(s,t)]^2\n
&& - 2 T_1(s,t) T_1^+(s,t) \cos(\th_\tau - \th) \n
&& +\ T_2^2(s,t) + [T_2^+(s,t)]^2 \n
&& -\ 2 T_2(s,t) T_2^+(s,t) \cos(\th_\tau + \phi_\tau - \th - \phi) \n
&& +\ 2 T_1(s,t) T_2(s,t) \cos(\phi) \n
&& +\ 2 T_1^+(s,t) T_2^+(s,t) \cos(\phi_\tau) \n
&& -\ 2 T_1^+(s,t) T_2(s,t) \cos(\th_\tau - \th - \phi) \n
&& -\ 2 T_1(s,t) T_2^+(s,t) \cos(\th - \th_\tau - \phi_\tau)\
\label{xsec_minus_independent}
\eqx
Similarly, taking the squared modulus for the proton-antiproton elastic scattering $|{\cal A}^{p\bar p}|^2 = |2{\cal A}_+ - {\cal A}|^2$ from Eq.~\eqref{appbar_independent} yields
\eq
\fr{d\sigma^{p\bar p}}{dt} &=& T_1^2(s,t) + 4[T_1^+(s,t)]^2 \n
&& - 4 T_1(s,t) T_1^+(s,t) \cos(\th_\tau - \th) \n
&& +\ T_2^2(s,t) + 4[T_2^+(s,t)]^2 \n
&& -\ 4 T_2(s,t) T_2^+(s,t) \cos(\th_\tau + \phi_\tau - \th - \phi) \n
&& +\ 2 T_1(s,t) T_2(s,t) \cos(\phi) \n
&& +\ 8 T_1^+(s,t) T_2^+(s,t) \cos(\phi_\tau) \n
&& -\ 4 T_1^+(s,t) T_2(s,t) \cos(\th_\tau - \th - \phi) \n
&& -\ 4 T_1(s,t) T_2^+(s,t) \cos(\th - \th_\tau - \phi_\tau)\ 
\label{xsec_appbar_independent}
\eqx
These relations provide a parameter-free extraction of the Odderon amplitude and, consequently, predictions for the $p\bar{p}$ elastic cross section. Both Eq.~\eqref{xsec_minus_independent} and  Eq.~\eqref{xsec_appbar_independent} inherit the parameters of FIT 1 and FIT 2 shown in Table~\ref{parameters}. At this point it is relevant to stress again  that FIT 1 does not fix the sign of $\phi$ since $\phi$ enters only through $\cos\phi$.  Both signs reproduce the TOTEM data equally well (hence the $\pm$ in Table~\ref{parameters}). This ambiguity does not exist for FIT 2 since $\phi_\tau$ depends explicitly on $\phi^+$. However, $\phi$ appears in Eqs.~\eqref{xsec_minus_independent} and~\eqref{xsec_appbar_independent} inside cosines whose arguments also involve $\theta$, $\theta_\tau$ and $\phi_\tau$, and not only through $\cos\phi$, the sign of $\phi$ does affect both cross sections. 

\subsection{5.4. Predictions for $p\bar p$ cross sections from scaling}

The $p\bar{p}$ elastic cross section predictions compared to the D0 measurement at $1.96$ TeV are given in Fig.~\ref{fig:d0_both}, left. We notice large differences between the different predictions, which can be understood from the sensitivity of the $p\bar{p}$ prediction to the underlying $pp$ fit, even though the fits to the TOTEM data are themselves very similar.

Two predictions for $p\bar{p}$ are shown: the orange curve (dip/bump region) and the blue curve (full domain). The blue curve lies far from the D0 data precisely because, in this case, the positive signature fit requires no negative signature (odderon) component, leading to an amplitude $\mathcal{A}_{-} = 0$. Such a prediction reduces to $\mathcal{A}^{p\bar{p}} = \mathcal{A}_{+}$, which does not reproduce the $pp$ dip and instead displays only a smooth shoulder, as shown by FIT 2  in Fig.~\ref{fig:fits} (right), thereby mimicking the $p\bar{p}$ behaviour. To make this explicit, we also show the green dotted curve (positive signature, full domain) in Fig.~\ref{fig:d0_both}, left, which illustrates how $|2\mathcal{A}_{+} - \mathcal{A}|^2$ overestimates the $p\bar{p}$ differential cross section. In contrast, the red curve (positive signature, dip/bump region) does produce a dip, showing that the odderon is required to consistently describe the $p\bar{p}$ cross section in the $|t|$ region covered by D0.

In Fig.~\ref{fig:d0_both}, right, we display the predictions for positive and negative values of $\phi$ for FIT 1, restricted to the dip/bump region only, since this shows the best compatibility with the $p\bar{p}$ elastic data (we recall that FIT 1 cannot determine the sign of $\phi$). A positive value of $\phi$ is clearly disfavored by the data, this is why we use negative values of $\phi$ in the following plots in this section.

\begin{figure*}[htbp]
    \centering
    \begin{minipage}{0.48\textwidth}
        \centering
        \includegraphics[width=\linewidth]{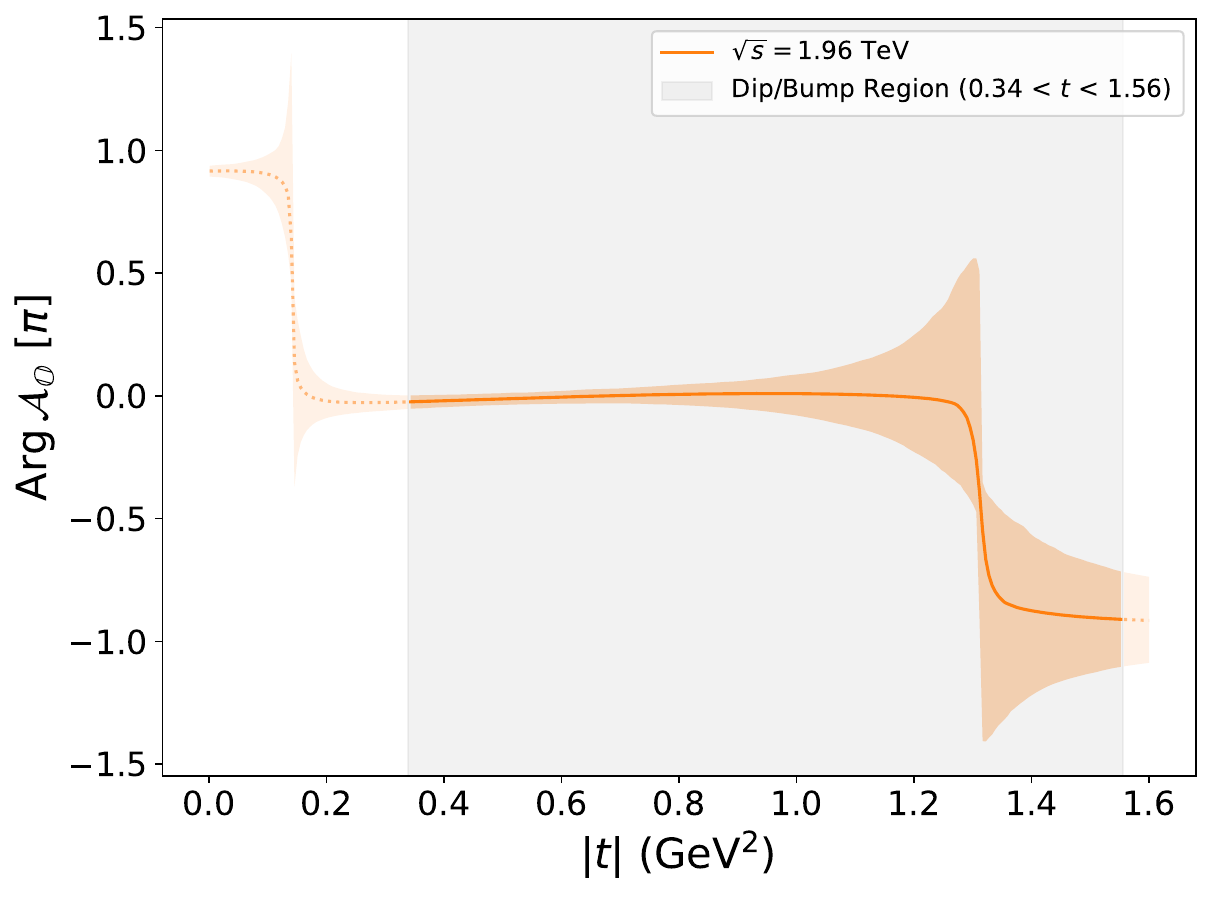}
    \end{minipage}%
    \hfill
    \begin{minipage}{0.48\textwidth}
        \centering
        \includegraphics[width=\linewidth]{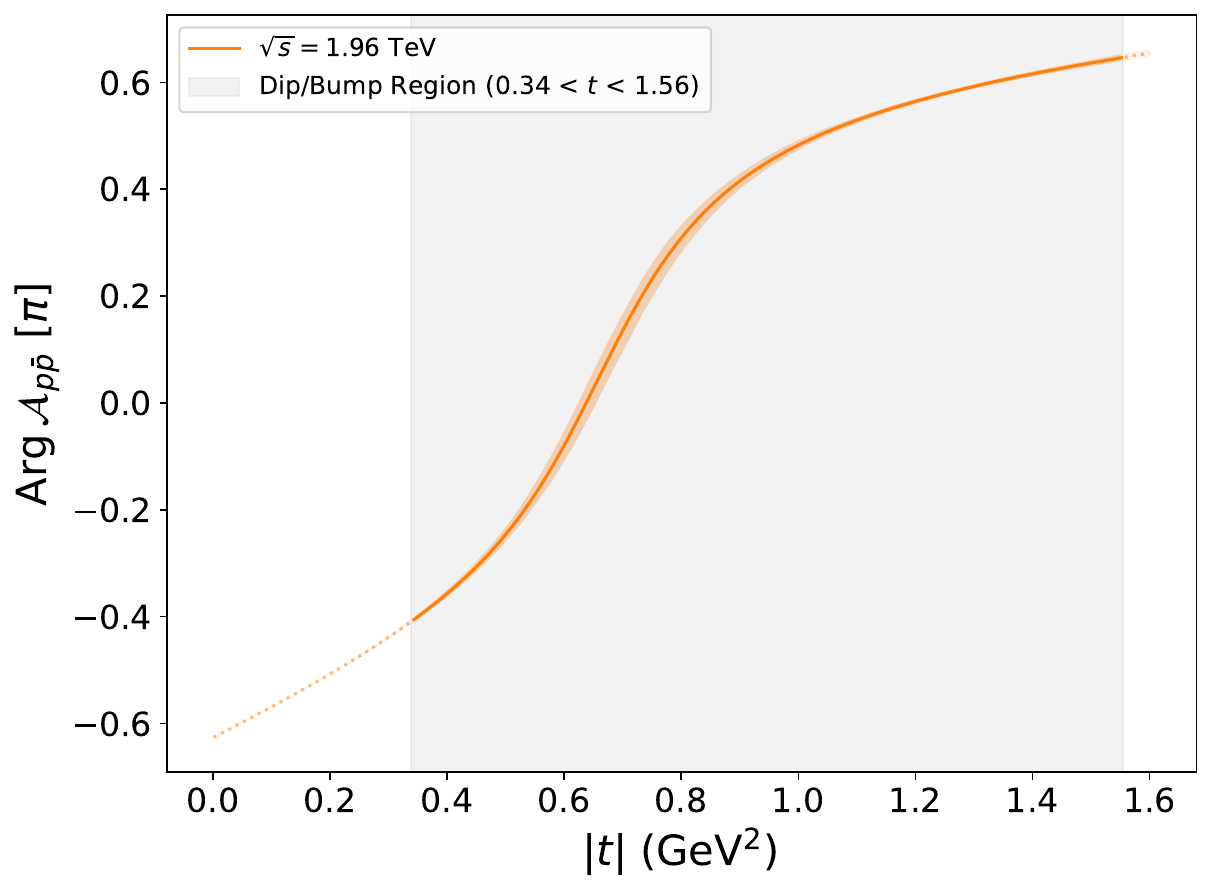}
    \end{minipage}

   \caption{\textbf{Left:} Prediction for the Odderon phase $\Psi = \mathrm{Arg}\,\mathcal{A}_{-}$ as a function of $|t|$ at $\sqrt{s} = 1.96$~TeV (D0 energy), derived from the TOTEM $pp$ fits via the scaling framework (negative-signature amplitude $\mathcal{A}_{-}$). \textbf{Right:} The corresponding prediction for the phase $\Psi_{p\bar p}$ of the elastic proton--antiproton amplitude, $\mathrm{Arg}\,\mathcal{A}_{p\bar{p}}$, obtained from the same fits through the combination     $\mathcal{A}_{p\bar{p}} = 2\mathcal{A}_{+} - \mathcal{A}$. The bands indicate the $1\sigma$ propagated uncertainties, and the TOTEM dip-bump region in the scaling framework is highlighted in grey.}
    \label{fig:d0_bothc}
\end{figure*}

\subsection{5.5. Amplitude and phases prediction for the Odderon and $p \bar{p}$}

The predictions for the odderon exchange differential cross section are given in Fig.~\ref{fig:d0_bothb} as a function of $t$ for $\sqrt{s} = 1.96$ TeV on the left and as a function of $t^{**}$ on the right, corresponding to the fit to the dip-bump region only. 

\begin{figure*}[htbp]
    \centering
    \begin{minipage}{0.48\textwidth}
        \centering
        \includegraphics[width=\linewidth]{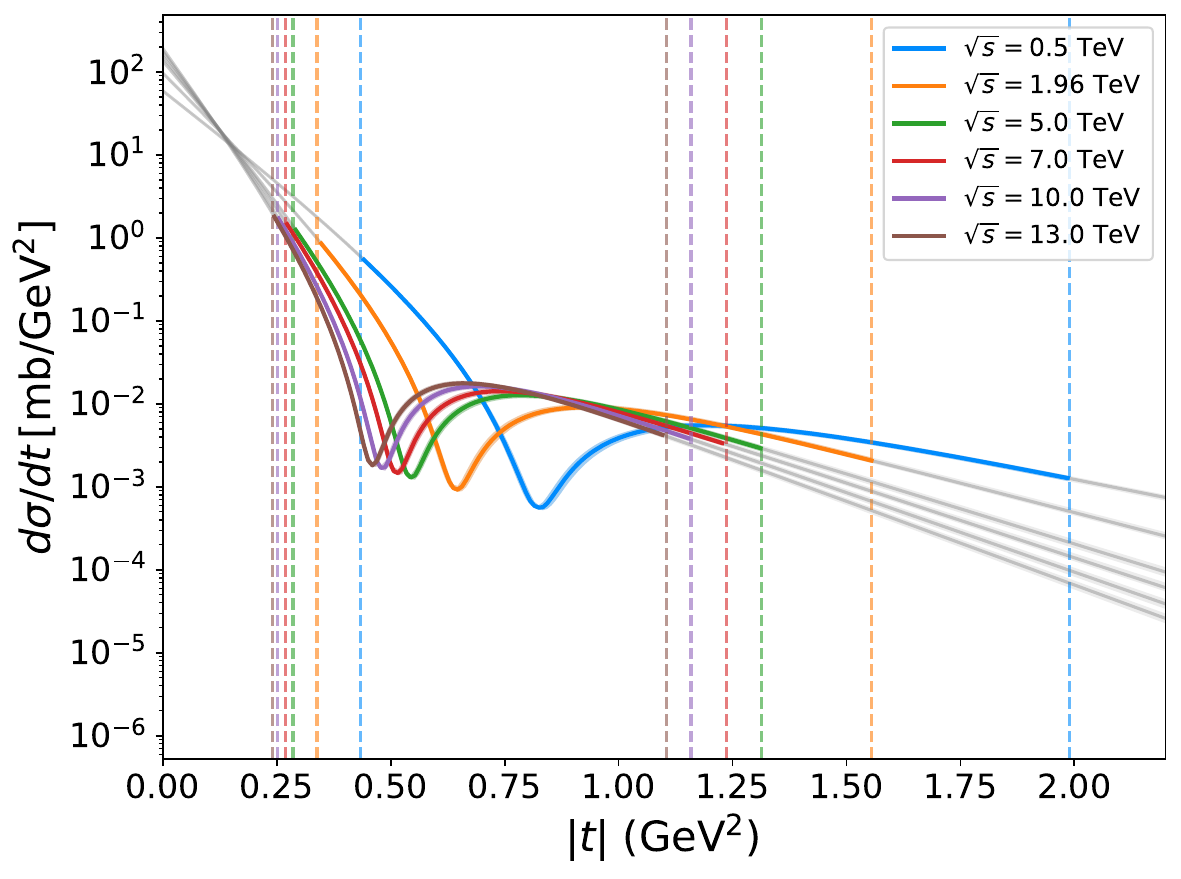}
    \end{minipage}%
    \hfill
    \begin{minipage}{0.48\textwidth}
        \centering
        \includegraphics[width=\linewidth]{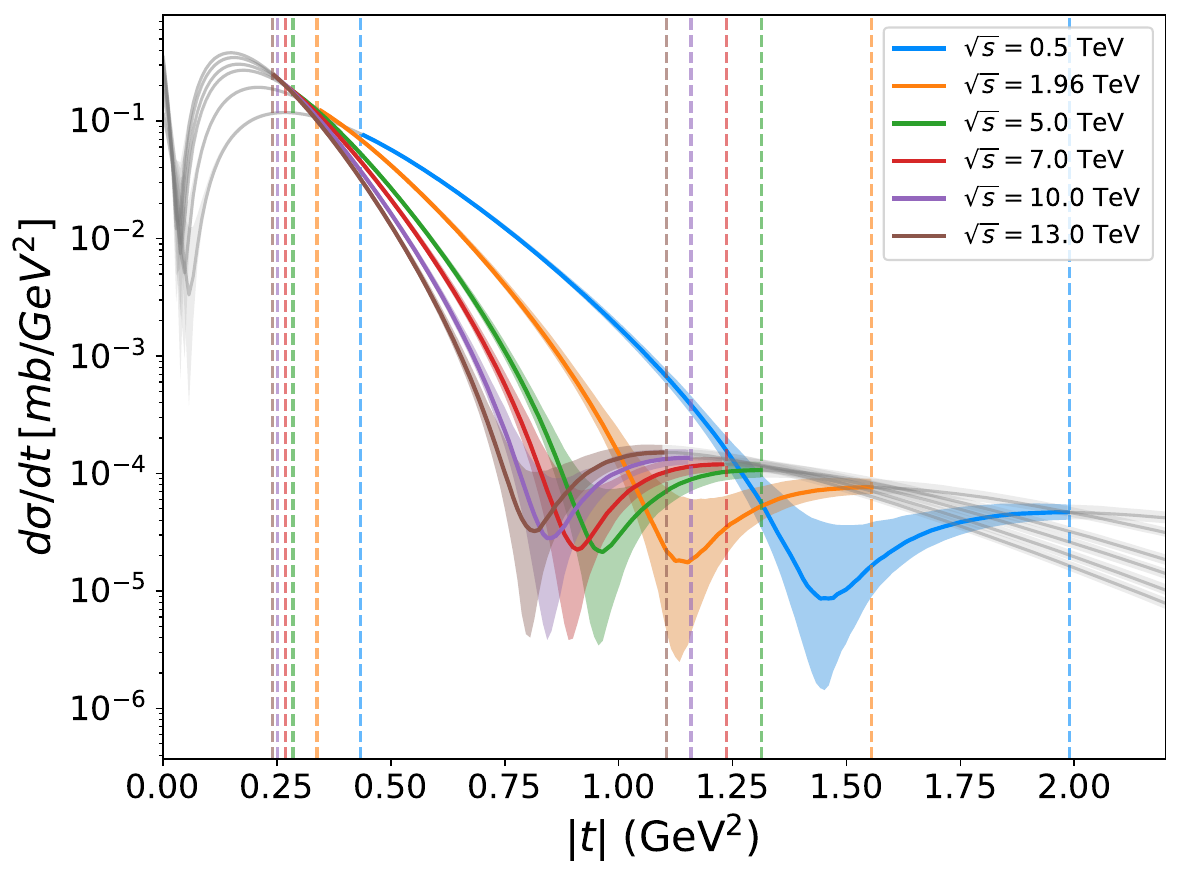}
    \end{minipage}

   \caption{\textbf{Left:} Predictions for the Pomeron signature on the elastic differential cross-section from $\sqrt{s} = 0.5$ TeV to $13$ TeV as a function of $|t|$, derived from TOTEM $pp$ fits via the scaling framework ($A_{+}$ amplitude). The shaded bands indicate the $1\sigma$ propagated uncertainties. The domains in $t$ corresponding to the dip bump region are indicated as dashed lines for each value of $\sqrt{s}$ (we recall that the region in $t^{**}$ is the same but it corresponds to different domains in $t$ for different $\sqrt{s}$). \textbf{Right:} The corresponding prediction for the Odderon signature.}
    \label{fig:d0_botpc}
\end{figure*}

From equations \eqref{aminus_independent} and 
\eqref{appbar_independent} one gets
\eq
{\rm Re}\,{\cal A}_-(s,t) &=& -\,N_1(s)e^{-B_1(s)|t|}\sin\th\;  \n
&& + N_1^+(s)e^{-B_1^+(s)|t|}\sin\th_\tau \n
&& -\,N_2(s)e^{-B_2(s)|t|}\sin(\th+\phi)\n
&&  \;+\; N_2^+(s)e^{-B_2^+(s)|t|}\sin(\th_\tau+\phi_\tau)
\label{ReAminus}
\eqx

\eq
{\rm Im}\,{\cal A}_-(s,t) &=& N_1(s)e^{-B_1(s)|t|}\cos\th \n
&&- N_1^+(s)e^{-B_1^+(s)|t|}\cos\th_\tau \n
&& +\,N_2(s)e^{-B_2(s)|t|}\cos(\th+\phi) \n
&& - N_2^+(s)e^{-B_2^+(s)|t|}\cos(\th_\tau+\phi_\tau)\ 
\label{ImAminus}
\eqx
and thus the Odderon amplitude phase $\Psi_{-}$
\eq
\Psi_{-} &=&\ {\arctan{\left({\rm Im}\,{\cal A}_-\over {\rm Re}\,
{\cal A}_-\right)}}\ .
\label{ArgAminus}
\eqx
In a similar manner, one writes
\eq 
\Psi_{p\bar p} &=&\ {\arctan{\left({\rm Im}\,{\cal A}_{p\bar p} \over {\rm Re}\,{\cal A}_{p\bar p}\right)}}\ . 
\label{ArgAppbar}
\eqx
The corresponding phase predictions are displayed in Fig.~\eqref{fig:d0_bothc}, using the dip/bump region and the  parameters from both fits that lead to a prediction in agreement with the $p\bar{p}$ data. Additionally, in Fig.~\ref{fig:d0_botpc}, we display the predictions for the Pomeron (left) and Odderon (right) $d\sigma/dt$ as a function of $|t|$ for different center-of-mass energies $\sqrt{s}$. The dip/bump region, translated from $t^{**}$ to $|t|$, is indicated by dashed vertical lines for each corresponding energy.

In Fig.~\ref{fig:d0_botpc}, we also display the predictions for the Pomeron (left) and the Odderon (right) $d\sigma /dt$ cross sections for different $\sqrt{s}$, as well as the dip-bump region in the fit displayed in dashed vertical lines. 
Each curve is displayed in grey (and not in different colors) outside the dip/bump region that corresponds to the fit to data.

\section{6.  Discussion}

Let us summarize the line of arguments which allows one  to  extract  the Odderon amplitude 
from the measurement of elastic cross-sections  at the LHC and the Tevatron.

The first step of the derivation is to notice that the Tevatron $ p\bar p$ elastic 
cross-sections do verify the same scaling found in the LHC $pp$ elastic cross-sections in
 the same domain of small nonzero momentum transfer, cf. Fig.~\eqref{fig333}.
From that property two consequences can be drawn:

 i) Scaling is a unified framework within 
 looking for the Odderon amplitude makes sense. 

ii) However, the Odderon contribution 
to cross-sections is expected to be small if not zero, due to the similarity of the $p\bar p$
 and $pp$ differential cross-sections in the scaling representation.
 
The second step of the derivation of the Odderon amplitude is to introduce a study of 
discrepancies shown by fits of $pp$ cross-secions using the positive signature  scaling amplitude $A_+.$ This is in contrast with the good fits obtained for the  amplitude $A_{pp}$ used in the original work \cite{scaling}.   The important point is that, if one finds such a difference, it indicates the existence of an Odderon contribution, since the Pomeron is of positive signature.

Such a discrepancy is indeed found if one restricts the fit analysis to the 
dip/bump range $ 0.6\lesssim t^{**} \lesssim 1.1\ \rm {Gev^2}$  where the scaling is osberved 
for $ p\bar p$ elastic cross-sections, indicated by the grey zone in Fig.~\eqref{fig:fits}.  One effectively finds a discrepancy with data in the depth of the dip, see Fig.~\eqref{fig:fits}, right. 
If, however, one keeps the full $ t^{**}$ range initially
 considered in the original work~\cite{scaling}, one obtains as expected a correct description in the region of the dip  near $t^{**} \sim 0.7,$ see Fig.~\eqref{fig:fits}, left. 

The third step is to confront  the predictions for the $p\bar p$ cross-sections coming from
 both cases: initial amplitude vs. positive signature, either in the full domain or in the 
 dip/bump ranges. Indeed extracting the negative signature amplitude allows one to make specific predictions for the  $p\bar p$ cross-sections. We are led to  consider the two options:
 
 i) The  $p\bar p$  prediction  using the difference between the scaling amplitudes $A_{pp}$ and $A_+$ found for the fit in the dip/bump $t^{**}$ range.
 
  ii) The $p\bar p$  prediction coming from the  fit for  $A_{pp}$ in the full  $t^{**}$ range. In this case the $A_{pp}$ amplitude seems to act as an effective positive signature, since it leads to a similar good prediction for both $ pp$ and $ p\bar p$ cross-sections. Apparently there is no need for an Odderon amplitude. However, as we shall see, this option is in contradiction with the phase prescriptions of the analytic S-matrix.
  
 Fig.~\eqref{fig:d0_both} shows that the two options lead to predictions in rough agreement with data,  taking into account that there are based only on higher energy $pp$ data.  However option ii) cannot be retained since the amplitude $ A_{pp}$ does not satisfy the energy-phase relation \eqref{relation} relevant for a positive signature amplitude featuring the equality between $pp $ and $p\bar p$ cross-sections. 

Other independent objections to option ii) are known. It has been shown that an Odderon contribution is necessary to explain the difference observed between the  $D_0$ data for $ p\bar p$ at  $\sqrt{s} = 1.96$ TeV and a precise extrapolation of the TOTEM cross-sections down to the same energy \cite{TOTEM:2020zzr}. A second argument comes from a study of the real over imaginary part of the amplitude near $t=0,$ where it is shown that a negative signature contribution is to be required. 

 So, we direct our attention to the option i), allowing us to extract a determination of the Odderon amplitude (both in modulus and phase) at high energies in the dip/bump region of momentum transfer.  The results are shown in Figs.~(\ref{fig:d0_bothb}, \ref{fig:d0_bothc}).

\section{ 7. Conclusion and outlook}
In this letter, we have shown that the combination of the phenomenological discovery of a scaling property of the $pp$ differential cross-section at the LHC~\cite{scaling} combined with the theoretical analyticity property of the energy-to-phase relation inherent to the S-matrix framework~\cite{chew} leads to a unique determination of  both a positive signature amplitude (cf. the Pomeron) and a negative signature one (cf. the Odderon). This last one is small but definitely nonzero, particularly in the dip region of the cross-section. It is as expected  with an exponentially decreasing  modulus in momentum transfer and an essentially zero phase, i.e., it is reall. 

As an outlook, one sees that the determination of the Odderon amplitude both in modulus and phase in a model-independent way provides the possibility to discuss the phenomenological Odderon  models of any kind
which have been proposed, possibly selecting those which  lead to a comparable amplitude. 

On a more general footing footing, it has been found since long~\cite{tolia} that new Regge  singularities of amplitudes could be found systematically near the so-called "wrong signature zeros" in the complex $l$-plane, see Eq.\eqref{regge}. This zero appears in this equation in the numerator at certain integer values of $l.$ Indeed, 
one of these zeros appears when the signature $ \eta = -1$ and the value $l=0$ ("intercept" $l+1$  in Regge language~\cite{chew}) near which stands precisely the  Odderon amplitude. This would relate the Odderon to a general mechanism whose candidates where discussed in Ref.~\cite{tolia}.

\section{Acknowdegments}
We thank C. Baldenegro for useful discussions about the work described in this paper.

\end{document}